\documentclass[journal]{IEEEtran}
\usepackage[T1]{fontenc}
\usepackage{cite}
\usepackage{amssymb}
\usepackage{amsfonts}
\usepackage{fancyhdr}  %
\usepackage{extarrows}
\usepackage{algorithm}
\usepackage{multirow,tabularx}
\usepackage{mathtools}
\usepackage{xcolor}
\usepackage{bm}
\usepackage{algorithmic}
\usepackage{amsmath}
\usepackage{subfigure}
\usepackage{makecell}
\usepackage{colortbl}
\usepackage{bbm}

\allowdisplaybreaks

\ifCLASSINFOpdf
  \usepackage[pdftex]{graphicx}
  \graphicspath{{../pdf/}{../jpeg/}}
  \DeclareGraphicsExtensions{.pdf,.jpeg,.png}
\else
  \usepackage[dvips]{graphicx}
  \graphicspath{{../eps/}}
  \DeclareGraphicsExtensions{.eps}
\fi
\begin{document}

\title{On Privacy, Security, and Trustworthiness in Distributed Wireless Large AI Models (WLAM)}
\author{       
    {Zhaohui Yang, Wei Xu, Le Liang, Yuanhao Cui, Zhijin Qin, and Merouane Debbah,~\IEEEmembership{Fellow,~IEEE }}
\thanks{Zhaohui~Yang is with the College of Information Science and Electronic Engineering, Zhejiang University, Hangzhou 310027, China, and the Key Laboratory of Information Processing, Communication and Networking of Zhejiang Province (IPCAN), Hangzhou 310007, China. (e-mail:  yang\_zhaohui@zju.edu.cn) }
\thanks{Wei Xu and Le Liang are with National Mobile Communications Research Laboratory, Southeast University,
Nanjing, China. (e-mails:  wxu@seu.edu.cn, lliang@seu.edu.cn) }
\thanks{Yuanhao Cui is with the School of Information and Communication Engineering, Beijing University of Posts and Communications,
Beijing, China. (e-mail: cuiyuanhao@bupt.edu.cn)}
\thanks{Zhijin Qin is with the Department of Electronic Engineering,
Tsinghua University,
Beijing, China. (e-mail: qinzhijin@tsinghua.edu.cn)}
\thanks{Merouane Debbah is with the Center for 6G Technology,
Khalifa University of Science and Technology,
Abu Dhabi, United Arab Emirates. (e-mail: merouane.debbah@ku.ac.ae)}
}

 \maketitle

\begin{abstract}
Combining wireless communication with large artificial intelligence (AI) models can open up a myriad of novel application scenarios. In sixth generation (6G) networks, ubiquitous communication and computing resources allow large AI models to serve democratic large AI models-related services to enable real-time applications like autonomous vehicles, smart cities, and Internet of Things (IoT) ecosystems. However, the security considerations and sustainable communication resources limit  the deployment of  large AI models over  distributed wireless networks. This paper provides a comprehensive overview of privacy, security, and trustworthy for distributed wireless large AI model (WLAM). In particular, a detailed privacy and security are  analysis for distributed WLAM is fist revealed. The classifications and theoretical findings about privacy and security in distributed WLAM are discussed. Then the trustworthy and ethics for implementing distributed WLAM are described. Finally, the comprehensive applications of distributed WLAM are presented in the context of electromagnetic signal processing.
\end{abstract}


\IEEEpeerreviewmaketitle

\section{Introduction}
\subsection{Motivation}
The integration  
of wireless communication and artificial intelligence (AI) technologies has revolutionized numerous sectors \cite{you2021towards}, from healthcare \cite{puri2024artificial} to transportation \cite{9181452}, by facilitating the development of distributed wireless large AI models (WLAM) \cite{10579546,10558824}. 
WLAM encompasses two aspects: large AI models deployed over wireless networks and large AI models designed to solve wireless communication problems.
Large AI models leverage the distributed nature of wireless networks to harness computational resources across diverse devices, enabling advanced data processing and real-time decision-making \cite{10558816}. 
However, the proliferation of WLAM introduces unique challenges, particularly in the areas of privacy, security, and trustworthiness.

As data is transmitted across wireless networks, it becomes susceptible to interception and unauthorized access, raising significant privacy concerns \cite{7879243}. 
To realize distributed WLAM, both federated learning and split learning are two potential driving technologies.  
In federated learning, for instance, while raw data remains on local devices, model parameters communicated to central servers can still be exploited by adversaries to infer sensitive information \cite{9069945}. Similarly, in split learning, the partial transmission of data between client and server networks creates potential vulnerabilities \cite{10121474}. These challenges necessitate robust encryption and privacy-preserving techniques to safeguard user data.

Moreover, the security of WLAM systems is paramount. Distributed models must be resilient against adversarial attacks that could compromise model integrity or manipulate outcomes. 
In WLAM, the model are is powerful and a large amount of data are obtained, which can introduce security concerns. 
Thus, it is required that models are not only secure but also transparent and accountable is essential for fostering user trust. In addition, the ethical deployment of WLAM involves addressing biases and ensuring fairness in AI-driven decisions \cite{ASHOK2022102433}, making trustworthiness a multifaceted challenge.

\subsection{Classifications of WLAM}
\begin{figure*}[t]
    \centering
    \includegraphics[width=\linewidth]{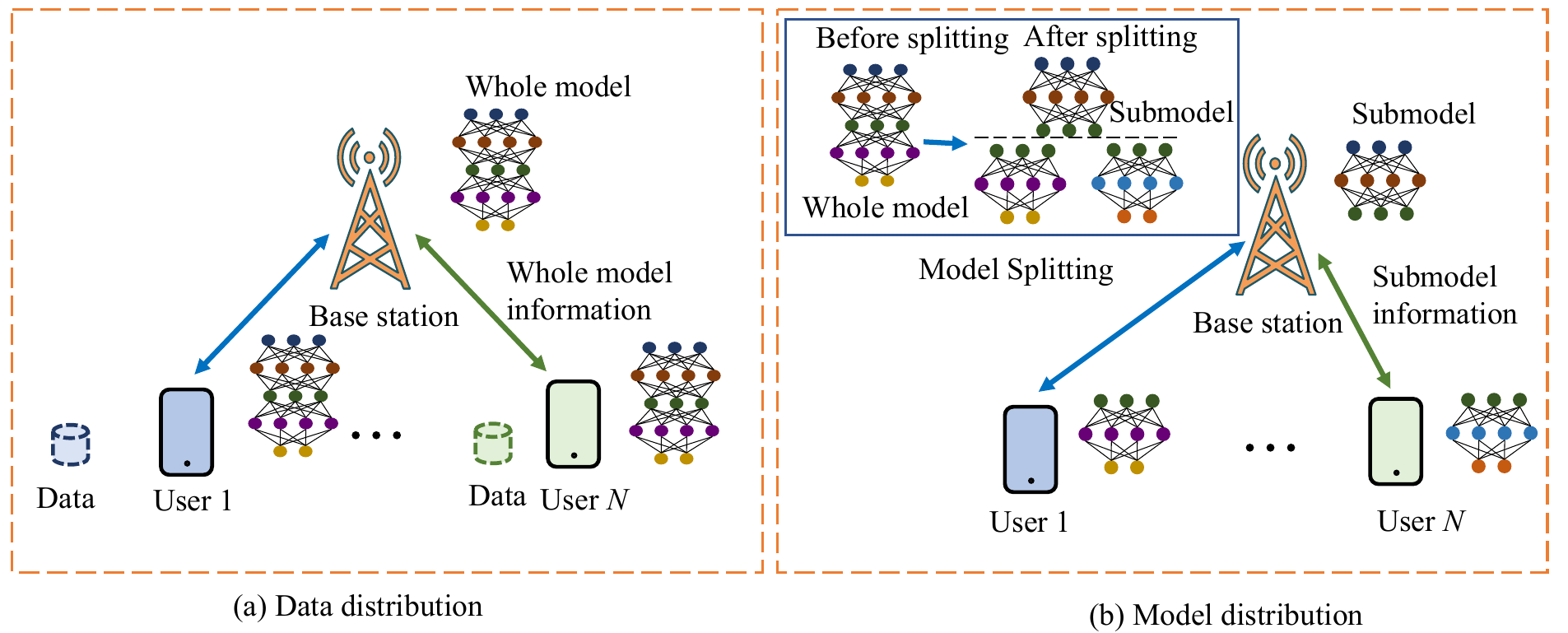}
    \caption{Two potential classifications of distributed WLAM.}
    \label{fig:DDMD}
\end{figure*}

WLAM can be broadly categorized into \textit{data distribution} and \textit{model distribution} approaches, as shown in Fig. \ref{fig:DDMD}, each offering distinct advantages and challenges.
\begin{itemize}
    \item \textbf{Data Distribution:} This approach involves distributing datasets across multiple devices. 
    Federated learning is a prime example, enabling devices to independently train local models and share updates with a central server
    \cite{9210812}. By transmitting only model updates rather than raw data, this method enhances privacy \cite{YANG202233}. However, it relies on sophisticated aggregation algorithms to ensure accurate global model convergence without leaking sensitive information. This balance between privacy and performance is a critical focus of ongoing research \cite{10283760}.
    \item \textbf{Model Distribution:} In contrast, model distribution involves dividing the large AI model itself across different devices. Split learning exemplifies this approach, where segments of the model are trained on separate devices \cite{Thapa_Mahawaga}. This method divides the original model into smaller sub-models, reducing the computational load on individual devices. However, model distribution method requires secure protocols for transmitting intermediate data and integrating model components, as partial data exchanges can expose vulnerabilities \cite{10.1145/3460120.3485259}. Ensuring the seamless integration of distributed model components remains a significant technical challenge.
\end{itemize}

Both data distribution and model distribution methods aim to maximize the efficiency of distributed resources while mitigating the inherent risks associated with wireless data transmission and model deployment.

\subsection{Contributions and Structure}
This paper provides an in-depth exploration of the privacy, security, and trustworthiness aspects of WLAM, offering insights into their implementation and challenges. Our contributions are threefold:
\begin{itemize}
    \item \textbf{Privacy Enhancement:} We explore a range of privacy-preserving techniques, including differential privacy, homomorphic encryption, and secure federated learning. These methods are critically evaluated for their effectiveness in protecting data while maintaining model accuracy. We propose the algorithms that incorporate noise addition and data obfuscation to enhance privacy without compromising performance. Our analysis highlights the trade-off between privacy protection and computational efficiency, providing guidelines for selecting appropriate strategies based on specific application needs.
    \item \textbf{Security Measures:} This paper examines advanced security strategies, including encryption algorithms and blockchain-based solutions. We discuss how blockchain technology can provide a decentralized framework for secure data management and model updates, ensuring integrity and traceability. Additionally, we explore the methods to defend against adversarial attacks and ensure model robustness. Our work highlights the importance of integrating security measures at multiple levels of the WLAM architecture, from data transmission to model deployment.
    \item \textbf{Trustworthiness and Ethics:} Addressing the ethical implications of WLAM deployment, we emphasize the importance of fairness, accountability, and transparency. We propose the frameworks for ethical analysis that incorporate fairness checks and explainable AI techniques, ensuring that WLAM systems operate within ethical boundaries. Our discussion includes case studies that illustrate the potential biases and ethical dilemmas inherent in AI-driven decision-making, along with strategies for mitigating these issues.
\end{itemize}

\begin{figure*}[t]
    \centering
    \includegraphics[width=\linewidth]{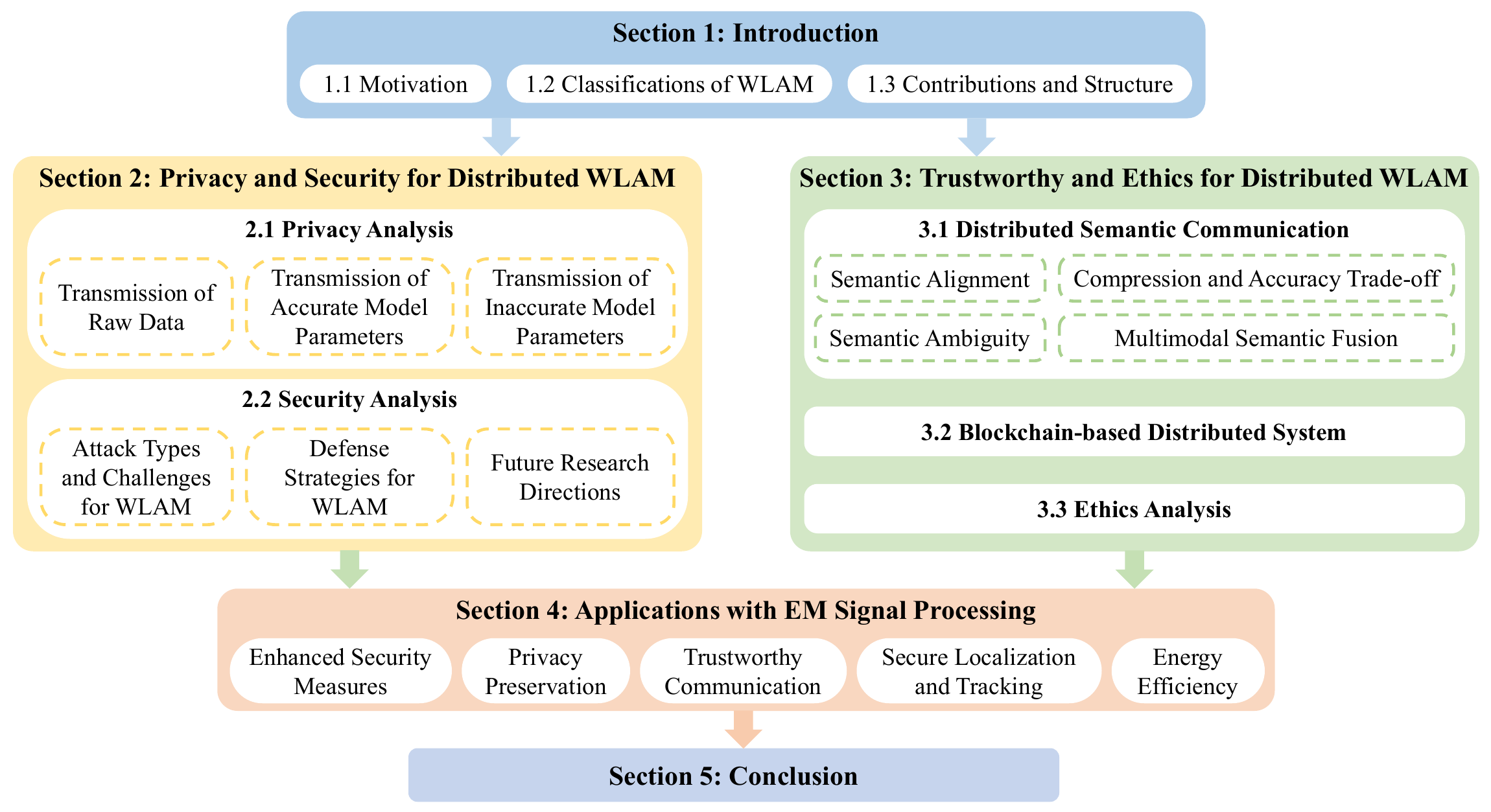}
    \caption{The structure of this paper.}
    \label{fig:structure}
\end{figure*}

The structure of this paper, as shown in Fig.~\ref{fig:structure}, is organized to facilitate a comprehensive understanding of these topics. Section \ref{sec.2} delves into privacy and security strategies for distributed WLAM, providing detailed analyses and case studies. Section \ref{sec.3} addresses the challenges of building trustworthy systems, exploring ethical considerations and proposing solutions. Section \ref{sec.4} concludes with a discussion on potential applications and future research directions, highlighting the importance of continued innovation in secure and ethical WLAM development.

\section{Privacy and Security for Distributed WLAM}\label{sec.2}
\subsection{Privacy Analysis}
In WLAM, model pre-training and fine-tuning require extensive data collection and processing.
Although original data can be stored locally, the training data transmitted over wireless communication in WLAM systems remains vulnerable to interception, potentially allowing attackers to recover the original data and compromise privacy.
The distributed WLAM system involves significant data interactions between devices and servers, increasing the risk of privacy breaches.
For example, in federated learning, local model parameters are uploaded from the client to the server \cite{9264742}, and in split learning, partially processed data is transmitted from the client network to the main server-side network, both of which are vulnerable to privacy risks.

To protect data privacy of the users during model training, encryption of transmitted information can prevent attackers from accessing the original data. This encryption can be categorized into three levels, raw data, accurate model parameters,  and inaccurate model parameters. 

\paragraph{Transmission of Raw Data}
The most basic privacy protection measure is to use privacy protection techniques such as physical layer security algorithms including artificial noise based secure beamforming 
to safeguard raw data transmitted between distributed users and servers. While this approach relies on the accurate channel gains,  
it offers limited privacy protection, making it possible for attackers to access raw data directly.

\paragraph{Transmission of Accurate Model Parameters}
Beyond direct data transmission, model parameters can be transmitted to enhance privacy. Through aggregation algorithms, global model parameters can be generated from sample data distributed across distributed users, thereby preventing direct access to the original data. Some existing model aggregation algorithms achieve performance similar to those of centralized machine learning \cite{qu2023privacy}. Additionally, transmitting model parameter data requires less bandwidth than raw data, conserving communication resources and building an energy-efficient WLAM system. However, attackers may exploit model parameters by constructing neural networks to reconstruct the original data, thereby posing a risk of privacy leakage. To handle this issue, this paper proposes a personalized federated learning and privacy protection algorithm based on information fusion. Privacy protection is achieved by adding noise according to user preferences and retaining the final layer locally to enhance data confidentiality. The proposed algorithm maintains a level of accuracy comparable to that of unencrypted cases while enhancing privacy.

\paragraph{Transmission of Inaccurate Model Parameters}
To prevent privacy breaches, original model parameters can be obfuscated by transmitting inaccurate model parameters. 
Due to transmitting inaccurate model parameters, the mapping of raw data and transmitted model parameters is converted from one-to-one to one-to-many or many-to-one.
Thus, it becomes more difficult for the  attackers to recover raw data. Typical inaccurate model parameters transformation methods include linear and nonlinear transformations.

For linear transformations, an example is adding artificial noise to model parameters:
\begin{equation}
    g_i (\nabla f_i (x)) = \nabla f_i (x) + n_i 
\end{equation} 
where \( x \) is the original data, \( \nabla f_i (x) \) is the gradient of the local model parameters to be transmitted, \( g_i (\nabla f_i (x)) \) is the transformed inaccurate gradient information, and \( n_i \) is the artificially added noise. The noise power increases linearly, whereas the gradient amplitude power grows quadratically. Therefore, when there are many users, an accurate global model can be obtained, allowing the federated learning process to converge effectively.

For nonlinear transformations, typical methods include quantized gradient, gradient encoding, and encryption algorithm.
\begin{itemize}
    \item \textbf{Quantized Gradient:} This method involves quantizing gradient information. One way of constructing the binary quantized gradient can be given by 
    \begin{equation}
        g_i (\nabla f_i (x)) = \begin{cases} 
    1, & \nabla f_i (x) \geq 0, \\
    -1, & \text{else}.
    \end{cases} 
    \end{equation}
   Convergence in model training is achieved by transferring only the gradient descent direction. The work in \cite{lang2023joint} uses random lattice-based vector quantization to add noise to transmitted parameters, avoiding common privacy leakage attacks without significantly affecting performance.
    \item \textbf{Gradient Encoding:} Some users upload only certain gradients, while others upload all gradients collectively. Differential privacy or gradient compression can mitigate gradient-based data reconstruction attacks to some extent but may impact accuracy. In \cite{liu2023privacy}, a privacy-preserving federated learning method called privacy-encoded FL is proposed, introducing an entangled nonlinear mapping between model gradients and original data by decomposing weight matrices into sub-matrices. This approach significantly reduces the risk of data reconstruction attacks without degrading performance. 
    \item \textbf{Encryption Algorithm:} To address privacy issues in WLAM, encryption techniques  like differential privacy, multi-party computation, homomorphic encryption, and functional encryption are used. 
    Differential privacy methods can be employed to ensure that individual data cannot be inferred from the aggregate outputs of the model. Homomorphic encryption, on the other hand, enables computation on encrypted data, eliminating the need to share raw data with servers.
    
    In \cite{hijazi2023secure}, a secure federated learning approach is implemented by combining fully homomorphic encryption with FL in IoT-enabled smart cities, achieving strong securityalong with high accuracy, low communication overhead, and reduced latency.
\end{itemize}

Inaccurate data transmission may hinder learning convergence, while accurate transmission poses privacy risks. Optimizing privacy protection measures by combining communication and computation processes to balance privacy enhancement and convergence speed is a valuable research direction.

\subsection{Security Analysis}

The security of distributed WLAM is crucial. Due to the distributed and dynamic nature of wireless communication networks, WLAM systems are vulnerable to various attacks, which can compromise model integrity, manipulate outputs, or disrupt operations.
In this subsection, we discuss the primary security threats faced by WLAM, including model injection attacks, Byzantine attacks, and other adversarial interventions. We also propose innovative countermeasures and highlight future research directions.

\subsubsection{Attack Types and Challenges for WLAM}

\paragraph{Data Poisoning Attacks}
For example, data poisoning attacks, in which malicious actors inject false data to mislead the AI model, pose a significant threat to system reliability.
A key countermeasure against such attacks involves implementing robust data validation mechanisms to identify and isolate malicious contributions in WLAM.

\paragraph{Model Injection Attacks and Environmental Complexity}

In wireless distributed scenarios, the complexity of model injection attacks increases significantly due to the dynamic environment. Attackers can exploit the instability of wireless networks or communication delays between nodes to inject malicious data or models, affecting the performance of the global model. Unlike traditional centralized systems, WLAM faces more complex channel conditions, making it challenging to detect malicious injections. For instance, attackers may hijack edge devices and leverage the interference characteristics of multipath channels to inject small perturbations at different intervals, leading to a gradual degradation of model performance, which is difficult to detect. Such progressive injection attacks are particularly covert in real-time applications.

\paragraph{Dynamic Byzantine Attacks and Adaptive Strategies}

Traditional Byzantine attacks assume that malicious nodes behave consistently throughout the training process. However, in WLAM, attackers can adopt \textit{dynamic Byzantine strategies}, where malicious nodes randomly switch between normal and malicious behavior during different training rounds, making attack detection more difficult. To counter such dynamic attacks, a \textit{dynamic trust evaluation mechanism} can be used, which models the historical behavior of each node probabilistically and employs time series analysis to evaluate node trustworthiness in real-time, thereby better identifying Byzantine nodes with dynamic behavior.

\paragraph{Channel Manipulation Adversarial Attacks}

The nature of wireless channels makes adversarial attacks more complex. Attackers can manipulate channel conditions (e.g., gain or fading) to interfere with the training and inference processes of the model. Compared to fixed adversarial perturbations, the uncertainty of wireless channels makes models more vulnerable to channel-level interference. Introducing \textit{channel-aware adversarial training}, where channel state information is included during training, can enhance robustness against channel interference. Additionally, leveraging \textit{multipath transmission and channel randomization} techniques can effectively reduce the attacker's control over the channel, thus improving defense against adversarial attacks.

\subsubsection{Defense Strategies for WLAM}

\paragraph{Cross-layer Aggregation Optimization}

To counter Byzantine attacks, a \textit{cross-layer collaborative aggregation mechanism} is proposed, which combines physical layer channel information and application layer model parameters for dual verification. During model parameter aggregation, anomalies in parameter values are considered alongside channel quality indicators to determine node trustworthiness, thereby improving the security of the aggregation process.

\paragraph{Intelligent Aggregator}

An \textit{intelligent aggregator} based on large models is introduced, which uses deep learning techniques to automatically detect abnormal parameter updates. By learning the update patterns of normal nodes, it can identify potential malicious updates. The intelligent aggregator continuously optimizes throughout the training process, enhancing adaptability to attacks.

\paragraph{Adversarial Game Modeling}

The interaction between attackers and defenders can be modeled as a game using the \textit{Stackelberg game model} from game theory. The defender acts as the leader, predicting the attacker's optimal strategy and making corresponding defense adjustments. In this way, defense strategies can be dynamically adjusted to minimize the impact of potential attacks.

\paragraph{Incentive Mechanism Design}

An incentive mechanism based on game theory is designed to reward honest behavior and penalize malicious behavior, encouraging nodes to follow the protocol and reducing the likelihood of attacks. Considering the resource-constrained nature of wireless networks, a lightweight incentive mechanism can be implemented to ensure practicality.

\paragraph{Collaborative Security Protection Using Edge Computing}

In the context of WLAM, collaboration among edge devices is crucial for enhancing system security. A \textit{collaborative edge protection mechanism} is proposed, where edge nodes work together to detect and share attack information, enabling distributed detection and response. Leveraging the \textit{heterogeneity of edge nodes} can improve detection accuracy for different types of attacks.

\paragraph{Channel Encryption and Privacy Protection}

By integrating \textit{channel encryption} and \textit{differential privacy} in edge computing, data confidentiality during transmission and node privacy can be ensured. This approach effectively prevents man-in-the-middle attacks and reduces privacy-related security risks.

\subsubsection{Future Research Directions}

\paragraph{Intelligent Adaptive Security Framework}

Develop an \textit{intelligent adaptive security framework} based on deep reinforcement learning, enabling the system to automatically learn the optimal defense strategy against different types of attacks. By continuously interacting with the environment, the framework can adjust defense measures in real-time to adapt to changing wireless environments and attack patterns.

\paragraph{Enhanced Federated Adversarial Training}

An \textit{enhanced federated adversarial training method} is proposed, which combines channel characteristics and adversarial sample generation. By dynamically generating adversarial samples that consider channel conditions during training, the model's robustness in wireless environments can be improved. This method can enhance defenses against adversarial attacks without significantly increasing computational overhead.

\paragraph{Large Model-driven Explainable Security Analysis}

Utilizing the powerful reasoning capabilities of large models, develop \textit{explainable security analysis tools} to monitor and interpret the decision-making process of models in real-time, helping to identify potential attack behaviors. By visualizing the internal state of the model, understanding of attack paths can be improved, thereby enhancing the model's response to security threats.

\paragraph{Blockchain-based Trust Management System}

Introduce \textit{lightweight blockchain technology} combined with edge computing nodes to build an efficient trust management system. The immutability and transparency of blockchain ensure that the behavior of each node is reliably recorded, thereby enhancing the overall trustworthiness of the system. Additionally, \textit{smart contracts} can be used to automate node behavior auditing and incentive distribution.

\subsubsection{Summary}

In WLAM systems, security is a key factor for successful deployment. Compared to traditional methods, innovative countermeasures such as \textit{cross-layer collaborative protection}, \textit{game theory-based adaptive defense}, and \textit{collaborative security mechanisms using edge computing} can more effectively address the unique challenges in wireless environments. In the future, intelligent adaptive security frameworks, enhanced adversarial training, and blockchain-based trust management are expected to improve WLAM security and robustness, providing a solid foundation for their practical deployment.

\section{Trustworthy and Ethics for Distributed WLAM}\label{sec.3}


In the landscape of distributed WLAM, it is essential to achieve trustworthiness and ethical standards for secure, reliable, and fair deployment. This section explores trustworthiness and ethical considerations in distributed semantic communication systems, blockchain-based distributed systems, and concludes with an ethical analysis addressing key challenges unique to WLAM.

\subsection{Distributed Semantic Communication}

By transmitting the context and intent behind the source data, referred to as semantic information, semantic communication enhances the data transmission efficiency \cite{Guler2018semantic}.
Given the intractable challenges associated with identifying semantic information for various communication tasks, advanced learning algorithms and model designs—especially large AI models—have been applied to semantic communication, facilitating semantic extraction, transmission, and interpretation \cite{ZHAO2024107055,huang2023toward,xing2024representation}. Thanks to the escalating computational capabilities of edge devices, they can now easily  support and execute learning-based models. By processing semantic extraction locally, the amount of data to be transmitted is reduced, promoting energy-efficient and low-latency communication \cite{dong2023semantic,ding2024joint,10550151}.

\begin{figure*}[t]
    \centering
    \includegraphics[width=\linewidth]{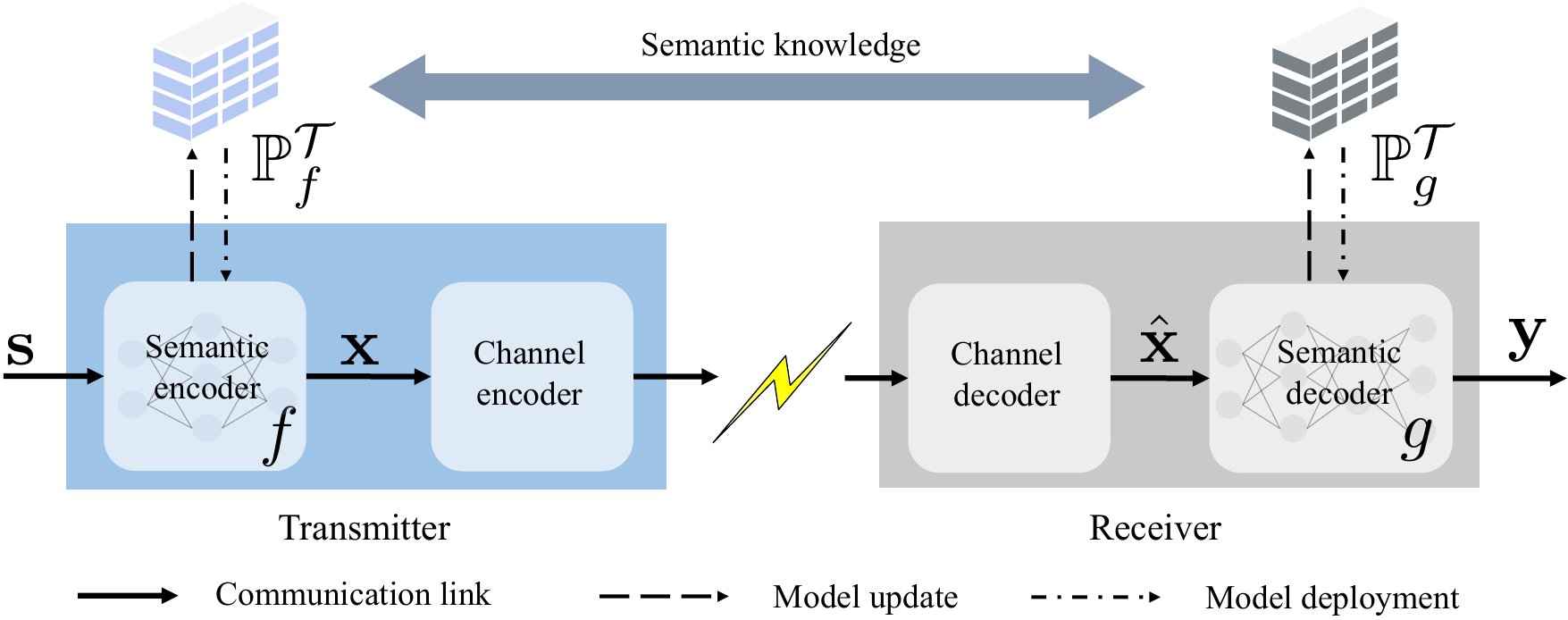}
    \caption{Illustration of model distributed semantic communication system.} 
    \label{fig:semcom}
\end{figure*}

The general framework of a distributed semantic communication system is presented in Fig.~\ref{fig:semcom}. The deep learning models, empowered by semantic knowledge, are distributed across the transmitter and receiver, denoted by $f$ and $g$ with parameter sets $\mathbb P^\mathcal{T}_f$ and $\mathbb P^\mathcal{T}_g$ respectively. Semantic knowledge can be accumulated during model training for communication task $\mathcal{T}$, where the optimized model parameter sets, denoted by $\mathbb P^\mathcal{T}$, encapsulate task-specific knowledge and are distributed to the transmitter and receiver. Using these deep learning models, the transmitter encodes the extracted semantic information $\bf x$ from the sensed data $\bf s$, and the receiver interprets the recovered semantic information $\hat {\bf x}$ for task inference $\bf y$.

In distributed WLAM for semantic communication, trustworthiness is pivotal for accurate and secure transmission of semantic information between transmitter and receiver nodes. To this end, several critical challenges need to be meticulously addressed, arising from  the intrinsic characteristics of deep learning and accurate semantic representation of data. Particularly, significant challenges inherent to deep learning models include their tendency to exhibit degraded generalization performance when there is a misalignment of semantic knowledge, as well as the trade-off between computational overhead and communication performance during training and inference. Additionally, challenges related to the accurate semantic representation, including semantic ambiguity and  semantic integration of multimodal data, further complicate the deep learning-based semantic communication systems. Large AI models can significantly mitigate these issues by providing robust semantic understanding, efficient compression techniques, and advanced disambiguation and multimodal fusion capabilities, thus  improving overall communication performance. These challenges and utilization of large AI models are detailed as follows.


\subsubsection{Semantic Alignment}
Existing deep learning-based semantic communication designs typically build the semantic knowledge for distributed models by training on large-scale datasets containing rich and diverse information. These learning strategies, including end-to-end learning \cite{xie2021deep} and multi-stage learning \cite{jankowski2020wireless}, ensure effective semantic alignment for specific communication tasks. The distributed models learn together, share knowledge, and improve collectively.

However, deep learning models tailored for specific tasks often face challenges in generalization performance due to misalignment of semantic knowledge between the training and inference phases. This issue arises because models are typically optimized for the specific characteristics of the training data and may not adapt well to novel, unseen inference data with different attributes or distributions. We formalize this discrepancy as a change within the underlying task, expressed by
\begin{align}
    d =\mathbb E_{\mathbf s \sim p(\bf s)}  \mathcal{D}\Big(g(f(\mathbf{s};\mathbb P_f^{\mathcal{T}_{\rm pre}});\mathbb P_g^{\mathcal{T}_{\rm pre}}),g(f(\mathbf{s};\mathbb P_f^{\mathcal{T}_{\rm new}});\mathbb P_g^{\mathcal{T}_{\rm new}})\Big), 
\end{align}
where $d$ quantifies the discrepancy, $\mathcal{D(\cdot)}$ is a discrepancy measure function that evaluates the difference in semantic knowledge represented by the parameters, $\mathbb P^{\mathcal{T}_{\rm pre}}$ represents the parameter set optimized for the previous task $\mathcal{T}_{\rm pre}$, and $P^{\mathcal{T}_{\rm new}}$ signifies the parameter set optimized for the new task $\mathcal{T}_{\rm new}$. A large $d$ suggests a larger discrepancy and potential degradation in performance, while a small $d$ indicates greater alignment and potentially better generalization. 

In order to achieve semantic alignment, one direct solution is model retraining for each new task; however, this approach is time-consuming, resource-intensive, and impractical in dynamic environments where tasks frequently change. Techniques such as domain adaptation and transfer learning, as applied in \cite{zhang2022deep} and \cite{yang2022semantic}, mitigate these issues by adjusting the model to handle variations in data distribution. However, domain adaptation often requires access to both labeled source data and unlabeled or partially labeled target data, which may be difficult to obtain. Additionally, accurately estimating the shift between the source and target domains can be challenging, especially when the shift is subtle or complex. On the other hand, transfer learning can suffer from overfitting if the model is fine-tuned on a small target dataset, and the pre-trained model's representations may be ineffective if the source task differs too much  from the target task. Furthermore, the choice of layers to fine-tune and the learning rate used for fine-tuning can significantly impact the performance of the transferred model. 

As efficient solutions, large AI models trained on diverse datasets have the potential to capture a broader range of semantic knowledge,  enabling robust understanding and inference of semantic information across various contexts. These models, often referred to as foundation models, are pre-trained on large, varied datasets and can be fine-tuned with fewer resources to adapt to new tasks, thereby mitigating the challenges associated with semantic misalignment.

\subsubsection{Semantic Compression and Accuracy Trade-off}
In semantic communication, semantic compression facilitates the reduction of data size by removing redundant or less important information, while preserving the most meaningful parts. Semantic accuracy refers to the degree to which the compressed data represents the original semantic meaning. High semantic accuracy ensures that the information conveyed after compression closely matches the original meaning, minimizing information loss and maintaining communication quality. The trade-off between semantic compression and semantic accuracy can be  characterized by the following aspects:
\begin{itemize}
    \item \textbf{Compression Level v.s. Information Loss:} As the compression level increases, the amount of information lost also increases. This can lead to a decrease in semantic accuracy, as critical details necessary for the correct interpretation of the information are removed. In the process of semantic recovery, receiving more information means less needs to be inferred, resulting in lower computational energy requirements. However, if too much information is lost during compression, task inference becomes more challenging, potentially leading to errors.
    \item \textbf{Selection Space and Computational Complexity:} The process of selecting appropriate semantic information from a large selection space is computationally intensive \cite{yang2023energy,e26050394}. The higher the compression ratio, the more difficult it becomes to select the right information, thereby increasing the computational complexity.
\end{itemize}

To address this trade-off, information-theoretic formulations, optimization methods, and adaptive compression techniques are employed. On one hand, semantic compression and accuracy are characterized based on information entropy theory, such as the informativeness of the extracted semantic information and the mutual information between the recovered semantic information and task inference \cite{shao2021learning,xu2023edge}. Using information-theoretic measures, an information bottleneck (IB)-based optimization problem is formulated  to characterize  the minimal and sufficient information required for a given task\cite{tishby2000information}. Specifically, the optimization problems aim to find a compressed representation of the input data that maximally preserves relevant information for predicting the task, while minimizing the amount of irrelevant information. To optimize the IB-based problem, variational inference is typically used to provide a tractable lower bound.

On the other hand, adaptive compression techniques adjust the compression level based on the importance of the information and the context. These techniques apply varying  compression levels to different layers of semantic information. Leveraging the powerful ability of large AI models to learn complex representations, these models can efficiently encode and compress semantic information while retaining essential details, as proposed in semantic communication. For instance, context-aware compression dynamically adjusts the compression rate based on the importance of the transmitted information, leveraging the understanding of context by large AI models to optimize performance. In \cite{guo2023semantics}, a semantic importance-aware communication scheme is proposed to quantify the importance of each frame and enable priority-based communication based on semantic importance. Generative language models, such as ChatGPT, quantify semantic importance by statistically analyzing the key terms within a frame. In contrast, discriminative language models, such as BERT, define the importance of words/frames based on the semantic loss caused by missing words/frames. By conveying frame importance across  various layers, priority-based parameter configurations can be implemented across layers. 
In \cite{lee2024integrating}, during the source compression phase, tthe initialized weights of the pre-trained language model BART are utilized to integrate the encoder-decoder transformer into the end-to-end SIONNA link-level communication system. Specifically, the AI Source Encoder labels the sentence \( \mathbf s \) as tokens, maps them, and encodes them into a state. The vector quantizer then quantizes the continuous feature vector into discrete indices using the codebook \( Z \). The receiver, using the shared codebook \( Z \), reconstructs the sentence. By utilizing pre-trained weights for initialization, the AI Source Encoder reduces dependence on specific channel conditions and data distributions during training.

\subsubsection{Semantic Ambiguity}
Semantic ambiguity refers to situations where the intended meaning of the transmitted data is unclear or can be interpreted in multiple ways. In wireless communication, it encompasses both linguistic issues, such as polysemy and contextual misinterpretation \cite{shi2021semantic}, and technical issues, such as signal degradation and interference \cite{luo2022semantic,10551876}. Specifically, polysemy occurs when a word or phrase has multiple valid interpretations depending on the context in which it is used. This can be problematic when the context necessary to disambiguate the meaning is not properly  conveyed. Even if a word or phrase is unambiguous in itself, the lack of proper context can lead to misinterpretation. Beyond the linguistic aspect, wireless distortion and interference introduce additional challenges related to signal quality. Although these technical issues are not directly related to the linguistic content, they can exacerbate semantic ambiguity by making it more difficult to recover the intended meaning. Robust signal processing algorithms and designs have been widely studied to minimize transmission errors. Therefore, in this section, we will focus on the linguistic issues of semantic ambiguity.

Existing methods for resolving semantic ambiguity include disambiguation protocols or context-specific rules. Specifically, disambiguation protocols offer a structured approach to resolving semantic ambiguity by defining clear and standardized rules for interpreting information. However, overly complex protocols may be challenging to understand and implement, potentially leading to errors or misinterpretations. Context-specific rules offer a more adaptable approach to addressing  semantic ambiguity by tailoring interpretations to specific contexts. These rules can be customized to fit the specific needs and conditions of a scenario, thereby improving the accuracy of interpretation. However, the effectiveness of these rules depends heavily on the accuracy and completeness of contextual information. Fixed protocols or rules may lack the flexibility needed to adapt to rapidly changing or unpredictable situations. Furthermore, implementing and maintaining these protocols can require significant resources, including time and effort for training and documentation.

Large AI models can address semantic ambiguity through their context-aware processing capabilities, which allow them to disambiguate information by considering the broader context in which the information is used. Hierarchical representations in deep neural networks capture complex relationships and contextual dependencies, enabling the model to understand nuanced meanings. Attention mechanisms within these models further enhance their ability to focus on relevant parts of the input, reducing the likelihood of ambiguity. Pretrained models, like transformers such as BERT and GPT, learn generalizable patterns and relationships from large corpora that can aid in disambiguation. By leveraging context-aware processing and attention mechanisms, large AI models can significantly mitigate the challenges posed by semantic ambiguity, enhancing  the accuracy and reliability of communication in both static and dynamic environments.
 
\subsubsection{Multimodal Semantic Fusion}
Fusing multimodal data in advanced applications, such as metaverse, presents  a significant challenge due to the heterogeneous nature of these data types~\cite{zhang2024unified}. Existing multimodal semantic communication designs not only explore the correlation among modaldata~\cite{zhu2024multi,xing2024representation}, but also consider the influence of the wireless channel for information fusion~\cite{luo2022multimodal}. Through multimodal data fusion, these designs enhance both data transmission efficiency and task inference performance. However, there are still ongoing challenges related to real-time processing, dependency on accurate contextual information, and the variability of data streams. Specially, the high computational demands of fusing and analyzing multiple data streams can strain the resources of wireless devices, especially those with limited processing power. The performance of multimodal semantic communication systems can degrade significantly if the contextual information is not precise or  if data is missing. Furthermore, variations in the quality or availability of different modalities may impact the effectiveness of the system. Addressing these challenges will be crucial for the continued advancement of multimodal semantic communication systems.

Large AI models, particularly those designed for multimodal processing, can effectively handle and fuse different data types, ensuring cohesive and accurate semantic interpretations~\cite{jiang2023large}. These models use shared latent spaces and advanced fusion techniques to seamlessly integrate information from various modalities. Cross-modal embeddings and techniques like cross-modal attention and multimodal fusion layers allow the model to learn joint representations that capture the relationships between different modalities. Adversarial training can ensure that the representations learned by the model are consistent across modalities. Advanced fusion techniques, such as late fusion, early fusion, and hybrid fusion, can integrate information from various modalities at different stages of the processing pipeline, leveraging the strengths of each modality while mitigating their weaknesses. By combining information from multiple modalities, large AI models can achieve a more comprehensive and accurate understanding of the underlying semantics, improving the robustness of the model against noise and missing data. This capability is essential for applications that require a holistic understanding of complex real-world scenarios, such as multimedia analysis, augmented reality, and multimodal dialogue systems.

\subsection{Blockchain-based Distributed System}
Blockchain-based distributed systems offer a promising framework for wireless communication networks, particularly for the management and optimization of resources in environments where trust, transparency, and security are essential. In the context of distributed WLAM, blockchain provides a decentralized infrastructure well-suited to handle the complex data and security requirements of these networks. Key elements of blockchain-based systems in wireless settings include nodes functioning as transmitters or receivers, distributed ledgers which record and verify information, consensus mechanisms like Proof of Stake or Practical Byzantine Fault Tolerance, and smart contracts like automated scripts that enforce rules on the blockchain. Together, these components create a self-sustaining network where nodes collaborate without central authorities, improving resilience, efficiency, and trustworthiness.

In a blockchain-based wireless network, data transmission flows from individual nodes to a shared blockchain ledger, where it is validated through consensus protocols and recorded immutably. This immutability ensures that once data is confirmed, it cannot be tampered with, thereby providing high levels of data authenticity and traceability. Blockchain's decentralized structure not only avoids single points of failure, which are common in centralized systems, but also offers a transparent and verifiable audit trail for every data interaction within the network. Privacy-preserving techniques, such as  zero-knowledge proofs, offer solutions for data validation without compromising sensitive information \cite{dhar2024blockchain}. Smart contracts further automate routine operations based on predefined rules, enabling consistent data handling across nodes without the risk of human error, thus enhancing the network's overall reliability.

Trustworthiness in blockchain-based wireless systems is crucial as it directly impacts the integrity and reliability of data used for decision-making and resource allocation in distributed WLAM. Given the dynamic and heterogeneous nature of wireless networks, data accuracy and consistency are vital, as erroneous or tampered data can disrupt real-time operations and impair model performance. Blockchain's decentralized and immutable framework helps ensure that every transaction and update within the system remains transparent, verifiable, and resistant to unauthorized manipulation, reinforcing user confidence. This trust is essential in privacy-sensitive scenarios, as blockchain's inherent transparency enables data validation without exposing user identities, making it an ideal solution for resource sharing and secure data handling in next-generation wireless networks\cite{hu2021blockchain}.

However, while blockchain offers significant benefits, maintaining trustworthiness in these systems presents several challenges. 
\begin{itemize}
    \item \textbf{Energy Consumption:} Traditional blockchain consensus protocols, like Proof of Work, are resource-intensive and may be impractical for wireless systems with power-constrained nodes, e.g., IoT devices. This limitation necessitates more energy-efficient consensus mechanisms suitable for wireless networks to maintain performance without exhausting device resources, as demonstrated by IoT implementations using the Raft consensus for increased throughput \cite{dhar2024blockchain}. 
    \item \textbf{Privacy Compliance:} Blockchain’s transparency can sometimes conflict with privacy needs, particularly in wireless applications where sensitive or location-specific data is involved. Immutable public ledgers may inadvertently expose user data, making compliance with privacy standards challenging. 
    \item \textbf{Security Risks:} Although blockchain architecture offers robust security, it is not immune to attacks. For instance, a 51\% attack, where a malicious actor gains control of the network's majority computational power, remains a significant threat, especially in wireless networks with heterogeneous nodes of varying security capabilities. The security advantages of blockchain are often tempered by this vulnerability, underscoring the need for complementary security measures within the system \cite{zhou2024ai}.
    \item \textbf{Scalability and Latency:} Blockchain’s consensus mechanisms, while reliable, often lead to slower data processing times. This delay can become a bottleneck as network traffic grows, which is problematic in real-time wireless applications where rapid responses are critical. The issue is especially pertinent in large-scale networks with high device density \cite{zuo2023survey}.
\end{itemize}

To address these challenges, large AI models provide effective solutions that enhance blockchain's functionality and trustworthiness within wireless networks. First, large AI models can optimize consensus protocols by leveraging predictive algorithms to monitor and anticipate network activity, dynamically adjusting protocol parameters to minimize latency and enhance scalability. For instance, AI-enhanced dynamic resource sharing in 6G networks has shown significant improvements in adapting consensus processes to real-time conditions, effectively reducing processing delays without compromising reliability \cite{hu2021blockchain}. Second, AI-based security measures can proactively identify and mitigate threats by detecting abnormal network patterns, which is particularly useful for countering attacks. Machine learning models trained on diverse network data can autonomously monitor and flag potential security issues, thus strengthening the overall security of blockchain-based wireless systems \cite{zhou2024ai}. Third, AI models equipped with advanced data compression algorithms can significantly reduce the data loads on wireless networks, promoting scalability while maintaining semantic fidelity. Additionally, privacy-focused AI solutions, like federated learning, allow nodes to process data locally, transmitting only essential insights to the blockchain to maintain user privacy without sacrificing transparency, which is crucial in 5G and 6G wireless IoT networks \cite{zuo2023survey}. By integrating large AI models, blockchain-based distributed wireless systems can overcome critical issues related to scalability, security, and privacy, fostering a trustworthy, resilient, and efficient environment that meets the demands of next-generation wireless AI applications.

\subsection{Ethics Analysis}
In the context of WLAM, ethics center around issues of privacy, fairness, and accountability, given the sensitive nature of data in wireless networks and the extensive interconnectivity among nodes. The high data volume and sensitive information flow in wireless communication amplify these ethical considerations, making it essential to address privacy risks, prevent data misuse, and ensure that model outcomes are unbiased and transparent.

Ethical analysis in distributed WLAM typically involves technical methods, such as differential privacy for protecting user data, fairness checks to avoid biased outcomes, and explainable AI techniques to clarify decision-making processes. In semantic communication systems, ethical analysis prioritizes maintaining semantic consistency and transparency. For example, fairness checks can ensure that compression algorithms do not introduce bias or information loss that affects certain demographics disproportionately. Ensuring ethical compliance means preserving data accuracy during compression and anonymizing personal data during  transmission.  In blockchain-based distributed WLAM, ethical analysis emphasizes balancing transparency with privacy requirements. For instance, implementing privacy-preserving cryptographic protocols, such as zero-knowledge proofs, enables nodes to verify transactions without exposing personal information. Additionally, smart contracts must be reviewed and audited to ensure compliance with data protection standards, guaranteeing that automated processes uphold ethical principles. Through these measures, ethical analysis helps WLAM systems align technical performance with broader social values, fostering responsible and fair use across applications.

\section{Applications with EM Signal Processing}\label{sec.4}
The distributed nature of WLAM offers significant flexibility and robustness for electromagnetic (EM) signal processing.
For instance, WLAM can enable real-time beamforming by leveraging the computational power of edge devices to optimize wireless signal transmissions.
Moreover, WLAM supports dynamic spectrum management in wireless networks by predicting spectrum availability and efficiently allocating resources.

\begin{figure*}[t]
    \centering
    \includegraphics[width=\linewidth]{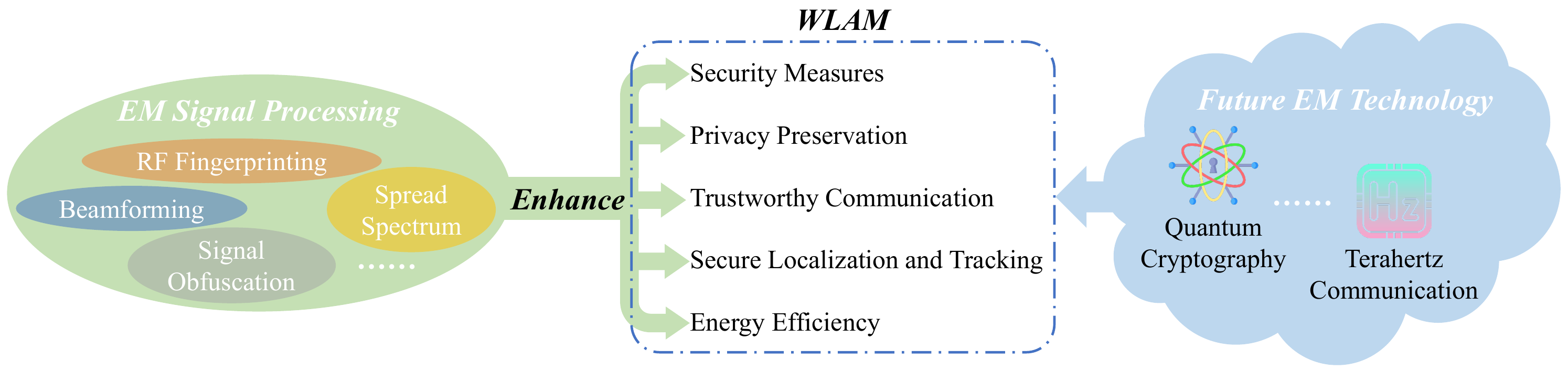}
    \caption{Application of EM signal processing in WLAM.}
    \label{fig:em}
\end{figure*}

Besides, EM signal processing \cite{stratton2007electromagnetic,paul2022introduction,van2007electromagnetic} can significantly enhance the privacy, security, and trustworthiness of distributed WLAM. By leveraging EM properties \cite{xu2023reconfiguring}, WLAM systems can achieve secure communication, robust privacy measures, and reliable system integrity, as illustrated in Fig.~\ref{fig:em}. This section delves into these aspects, highlighting their impact and potential advancements.

\begin{itemize}
    \item \textbf{Enhanced Security Measures:} EM signal processing techniques, such as physical layer security, exploit the inherent characteristics of EM waves to establish secure communication channels. These methods can effectively mitigate risks like eavesdropping and jamming by using techniques such as beamforming \cite{8030327} and spread spectrum \cite{10.1007/BFb0000435}. By dynamically adjusting signal parameters, WLAM systems ensure data integrity and confidentiality, creating robust defenses against potential security threats.
    \item \textbf{Privacy Preservation:} Privacy is a critical concern in WLAM systems, especially when dealing with sensitive data. EM signal processing can enhance privacy by implementing techniques like signal obfuscation and masking \cite{10.1145/3485832.3485894,10556606}. These methods obscure the data being transmitted, making it difficult for unauthorized entities to intercept and decipher the information. Additionally, the use of differential privacy at the signal level \cite{10.1145/2976749.2978318} ensures that individual data points remain protected while still allowing for accurate aggregate analysis.
    \item \textbf{Trustworthy Communication:} The trustworthiness of WLAM systems relies heavily on the reliability and authenticity of the data being transmitted. EM signal processing contributes to this by enabling secure key distribution and authentication protocols that ensure only authorized devices can communicate within the network. Techniques such as RF fingerprinting \cite{4211360} can verify the identity of devices based on their unique signal characteristics, preventing spoofing and unauthorized access.
    \item \textbf{Secure Localization and Tracking:} EM signal processing facilitates precise and secure localization and tracking, which are vital for applications requiring spatial awareness, such as logistics and augmented reality. By using methods like time-of-flight and RF fingerprinting, WLAM systems can accurately determine the position of devices while maintaining privacy. These techniques ensure that tracking data is securely transmitted and accessed only by authorized entities, safeguarding against potential misuse.
    \item \textbf{Energy Efficiency:} While enhancing security, EM signal processing also supports energy efficiency. Techniques such as adaptive modulation and coding adjust signal parameters based on current conditions, reducing power consumption without compromising security. This capability is essential for maintaining the sustainability of large-scale WLAM deployments, ensuring that security measures do not excessively drain device resources.
\end{itemize}

The future of WLAM systems will increasingly depend on advancements in EM signal processing to address evolving privacy and security challenges. Emerging technologies like terahertz communication and quantum cryptography offer promising avenues for further enhancing security protocols. These innovations have the potential to improve the speed, accuracy, and robustness of WLAM applications, ensuring that they remain secure and trustworthy as they adapt to new technological landscapes.
Further research is required to develop energy-efficient algorithms capable of operating seamlessly across diverse devices in distributed WLAM systems.

\section{Conclusions and Future Directions}
In this paper, we explored the critical aspects of privacy, security, and trustworthiness in distributed WLAM systems. As the integration of wireless communication and AI continues to advance, addressing these challenges becomes paramount for the successful deployment of WLAM systems. We examined various privacy-preserving techniques, including differential privacy and homomorphic encryption, highlighting their roles in protecting sensitive data. Additionally, we discussed security measures such as physical layer security and blockchain integration, which are essential in safeguarding against potential threats and ensuring data integrity. Furthermore, we emphasized the importance of trustworthiness, focusing on ethical considerations, transparency, and accountability in WLAM systems. By addressing biases and ensuring fairness, we can foster greater user trust and confidence in AI-driven decisions. Looking ahead, ongoing research and innovation in EM signal processing and AI will be crucial in overcoming the challenges associated with WLAM systems. By continuing to develop robust privacy, security, and trustworthiness frameworks, we can unlock the full potential of WLAM, paving the way for more intelligent, secure, and ethical wireless networks.

In conclusion, WLAM offers a promising framework for integrating AI with wireless communication systems, though it necessitates addressing inherent security and privacy challenges. 
Advancements in cryptographic techniques and distributed algorithms will be pivotal in overcoming these challenges and unlocking the full potential of WLAM.
Despite its potential, significant challenges remain for the widespread deployment of distributed WLAM. Scalability remains a challenge, particularly as the number of participating devices increases. Besides, ensuring low-latency communication among distributed components while maintaining data security is a critical challenge.
\bibliographystyle{IEEEtran}
\bibliography{IEEEabrv,ref,MMM}
\end{document}